\documentclass[onecolumn,showpacs,amsfonts,aps,prc,nofootinbib,floatfix,%
superscriptaddress]{revtex4}

\usepackage{amsmath}
\usepackage{bm}
\usepackage{graphicx}

\usepackage{epsfig}
\newcommand{\beq}{\begin{equation}}
\newcommand{\eeq}{\end{equation}}
\newcommand{\bea}{\vspace{0.25cm}\begin{eqnarray}}
\newcommand{\eea}{\end{eqnarray}}

\newcommand{\ro}{\mbox{{\boldmath
$\rho$}}}

\newcommand{\kb}{{{\bf k}}}

\newcommand{\bb}{{{\bf b}}}

\def\lsim{\mathrel{\rlap{\lower4pt\hbox{\hskip1pt$\sim$}}
    \raise1pt\hbox{$<$}}}         %less than or approx. symbol
\def\gsim{\mathrel{\rlap{\lower4pt\hbox{\hskip1pt$\sim$}}
    \raise1pt\hbox{$>$}}}         %greater than or approx. symbol

\newcommand{\landau}{L.D.~Landau Institute for Theoretical Physics,
        GSP-1, 117940, Kosygina Str. 2, 117334 Moscow, Russia}

\begin{document}

%%%%%%%%%%%%%%%%%%%%%%%%Front Matter%%%%%%%%%%%%%%%%%%%%%%%%%%%%%%%%%%
%%%%%%%%%%%%%%%%%%%%%%%%%%%%%%%%%%%%%%%%%%%%%%%%%%%%%%%%%%%%%%%%%%%%%%
%\renewcommand{\thefootnote}{\fnsymbol{footnote}}

\title{
Monte Carlo Glauber wounded nucleon model with meson cloud
}
\date{\today}

\author{B.G.~Zakharov}\affiliation{\landau}

\begin{abstract}
We study the effect of the nucleon meson cloud on
predictions of the Monte Carlo Glauber wounded nucleon model for $AA$, $pA$, 
and $pp$ collisions. 
From the analysis of the data on the charged
multiplicity density in $AA$ collisions we find that
the meson-baryon Fock component 
reduces the required fraction of binary collisions
by a factor of $\sim 2$ for Au+Au collisions at 
$\sqrt{s}=0.2$ TeV and  $\sim 1.5$ for Pb+Pb collisions at 
$\sqrt{s}=2.76$ TeV.
For central $AA$ collisions the meson cloud can increase 
the multiplicity density by $\sim 16-18$\%.
We give predictions for the midrapidity charged multiplicity density 
in Pb+Pb collisions at
$\sqrt{s}=5.02$ TeV for the future LHC run 2.
We find that the meson cloud has a weak effect 
on the centrality dependence of the ellipticity 
$\epsilon_2$ in $AA$ collisions. 
For collisions of the deformed uranium nuclei at $\sqrt{s}=0.2$ TeV
we find that the meson cloud may improve somewhat agreement with
the data on the dependence of the elliptic flow on the charged multiplicity
for very small centralities defined via the ZDCs signals. 
We find that the meson cloud may lead to a noticeable reduction of $\epsilon_2$
and the size of the fireball in $pA$ and $pp$ collisions.
  
\end{abstract}
%
%\pacs{12.38.Mh, 24.85.+p}

\maketitle

%%%%%%%%%%%%%%%%%%%OUR%%%%%%%%%%%%%%%%%%%%%%%%%%%%%%%%%%%%%%%%%%%%%%%
\section{Introduction}
Experiments at RHIC and LHC on heavy ion collisions 
give a variety of facts in favor of production of the hot QCD matter
in the quark-gluon plasma (QGP) phase.
Hydrodynamic analyses of the flow effects in $AA$ collisions at 
RHIC and LHC energies suggest that the QGP produced in $AA$ collisions
expands as a  near-ideal liquid \cite{Huov_hydro,Kodama_hydro}. 
The hydrodynamic simulations  support the production time
of the QGP $\tau_0\sim 0.5-1$ fm \cite{Heinz_hydro2,Heinz_tau}. 
However, a consistent  treatment of the QGP production 
is presently impossible. It does not allow to impose accurate 
from principles initial conditions for hydrodynamic simulations of the QGP
evolution in $AA$ collisions, and requires to use phenomenological models.

At present the most popular phenomenological methods in use for 
determination of the initial
conditions  for the plasma fireball
are the IP-Glasma model \cite{IP-GL1,IP-GL2} and the wounded nucleon Glauber 
model \cite{WNG,KN}. The IP-Glasma model is based on 
the pQCD color-glass condensate scheme \cite{CGC}.
It assumes that gluon fields of the colliding nuclei can be treated 
perturbatively down to an infrared scale $m_g\sim 1/R_p\sim 0.2$ GeV  
\cite{IP-GL1,IP-GL2} ($R_p$ is the proton charge radius). For such
a small infrared scale the gluon (and sea quark) density of the nucleon 
may be described as radiatively generated via the Weizs\"acker-Williams
fields of constituent quarks \cite{NZ12,Barone_EMC}.  
In this case the perturbative dipole cross section $\sigma_{q\bar{q}}$
of interaction of the $q\bar{q}$ pair with a nucleon (that can be
expressed via the gluon density \cite{NZ_peaks})
corresponding to the double gluon exchange  
allows to reproduce the $\pi p$ 
cross section in the tens of GeV region \cite{2G_Low}. 
However, the infrared cutoff scale $\sim 1/R_p$ is
in contradiction with the fact that 
the inverse gluon correlation radius in the QCD vacuum $1/R_c\sim 0.75$ GeV
\cite{Shuryak}. Because this scale, which is the natural lower limit for the
virtuality scale of the perturbative gluons, is 
several times bigger than $1/R_p$. One can expect that for the $q\bar{q}$
pair with the transverse size $\rho\sim R_p$ the inelastic interactions
are dominated by the nonperturbative process of the color flux tube
rearrangement \cite{Nussinov_2G,Nussinov_an}, and the perturbative mechanism
becomes dominating only at $\rho\lsim R_c$.  In the dipole approach 
 to the BFKL equation
\cite{NZ_gBFKL} 
the data on the low-$x$ proton structure function 
$F_2$ can be well described assuming that the dipole cross section contains 
the perturbative component with the infrared cutoff $m_g\sim 1/R_c$ and 
an energy independent nonperturbative component that can be fitted 
to reproduce experimental pion-proton cross section \cite{NZ_HERA}.     
It is important that in this scenario the BFKL evolution and the 
saturation effects in the perturbative dipole cross section with the infrared 
cutoff $m_g\sim 1/R_c$ turn out to be considerably weaker than that 
for $m_g\sim 1/R_p$. 
The recent analysis \cite{Zoller2} of the $pp$ cross section
in the above two component dipole scheme with $m_g \sim 0.75$ GeV shows that 
the perturbative contribution turns out to be smaller than the 
nonperturbative one up to $\sqrt{s}\sim 10^3$ GeV \cite{Zoller2}. 
This makes questionable the accuracy of the purely perturbative schemes 
for calculations of the initial QGP parameters in $AA$ collisions at RHIC
and LHC energies.

The wounded nucleon Glauber model \cite{WNG,KN} is a 
phenomenological extension of the ordinary Glaber model
invented for calculations of the hadron spectra in inelastic nucleus-nucleus
and hadron-nucleus collisions. 
In its original form \cite{WNG} it was assumed that in the $AA$ 
collision each nucleon undergoing inelastic soft interaction
(so-called participant or wounded nucleon) produces a fixed contribution to the 
multiplicity rapidity density. 
At first this idea was purely empirical. But recently there was an attempt
to give a QCD interpretation to this picture 
\cite{Bialas_brem1,Bialas_brem2,Bzdak_brem}. 
For particle production at midrapidity  ($\eta=0$) in 
the center-of-mass (c.m.) frame  of the colliding nuclei
the contribution of each wounded nucleon equals half of the $pp$ multiplicity
rapidity density. 
It gives for $AA$ collisions 
the multiplicity density $\propto N_{part}$, where $N_{part}$ is the number
of participants in both the colliding nuclei.
Later, in \cite{KN} it has been  proposed a two component version 
of the model that, in addition to the term $\propto N_{part}$, accounts 
for the hard binary collision term $\propto N_{coll}$. 
In this two component version the midrapidity 
multiplicity density in 
$AA$ collisions takes the form
\beq
\frac{dN_{ch}(AA)}{d\eta}=\frac{(1-\alpha)}{2}n_{pp}N_{part}+
\alpha n_{pp}N_{coll}\,,
\label{eq:10}
\eeq
where $n_{pp}=dN_{ch}/d\eta$ is the midrapidity charged multiplicity 
density in $pp$ collisions,
and $\alpha$ characterizes the fraction of hard processes
to multiparticle production. 
In the Glauber model 
model $N_{part}$ and $N_{coll}$ can be expressed through the inelastic 
$pp$ cross section $\sigma_{in}^{NN}$ and the nuclear density $\rho_A$. 
For the A+B collision
of heavy nuclei at a given impact parameter $\bb$ 
in the optical approximation they read
\bea
N_{part}(\bb)=\int d\ro T_{A}(\ro)\left\{1-
\exp[-T_B(\bb-\ro)\sigma_{in}^{NN}]\right\}
+\int d\ro T_{B}(\ro)\left\{1-
\exp[-T_A(\bb-\ro)\sigma_{in}^{NN}]\right\}\,,
\label{eq:20}
\eea
\beq
N_{coll}(\bb)=\sigma_{in}^{NN}\int d\ro T_A(\ro)T_B(\bb-\ro)\,,
\label{eq:30}
\eeq
where $T_A(\bb)=\int dz\rho_A(\bb,z)$ is the nuclear
profile function.

It is important that the two component Glauber model allows the 
Monte Carlo formulation \cite{PHOBOS_MC,GLISS,GLISS2}. The Monte Carlo Glauber 
(MCG) model has proved to be a useful tool for analysis of the event-by-event 
fluctuations of observables in $AA$ collisions.
Fitting data on centrality dependence of the charged particle multiplicity
density in Au+Au collisions at $\sqrt{s}=0.2$ TeV and in Pb+Pb collisions
at $\sqrt{s}=2.76$ TeV gives $\alpha\approx 0.13-0.15$
\cite{PHOBOS0405,STAR1,LHC_MCGL}. For such a value of $\alpha$ the hard
contribution to the particle production in $AA$ collisions turns out to 
be rather  
large ($\sim 40-50$\% for central collisions). 
%
%***************************
However, a considerable contribution of the binary collisions
was recently  questioned by the absence of the knee-like structure
in the STAR data on the flow coefficient $v_2$ in U+U collisions at 
$\sqrt{s}=193$ GeV \cite{UU_STAR} predicted in the MCG simulations of 
\cite{UU_Lednicky,UU_Voloshin}.  
In \cite{UU_Lednicky,UU_Voloshin} it was predicted 
that due to a prolate shape of the uranium nucleus
the initial $\epsilon_2$ should have a 
knee-like structure at multiplicities
in the top 1\% U+U collisions. The effect is related to the growth of 
the contribution of the binary collisions $\propto N_{coll}$ 
for the tip-tip configurations 
of the colliding nuclei as compared to the body-body events. 
The knee in the elliptic flow in U+U collisions  predicted in 
\cite{UU_Lednicky,UU_Voloshin} 
stimulated searches for different prescriptions for the entropy distribution
in $AA$ collisions in the Glauber picture that may be consistent
with a smaller (or without) contribution of the binary collision 
term \cite{UU_Bass,UU_Singh}.

The required contribution of the binary collision term may be smaller 
if the Glauber wounded nucleon model is formulated in the sub-nucleon level,
when inelastic $NN$ interactions are described as inelastic interactions
of the nucleon constituents, say, quarks. The formulation of the wounded
nucleon model at the quark level has been given in Refs. \cite{woun_q1,woun_q2}.
For the wounded nucleon models with internal sub-nucleon degrees of freedom  
the $dN_{ch}(AA)/d\eta$ is a nonlinear function of the number of the wounded 
nucleons even without the hard scattering contribution
\cite{Voloshin_3q,Bozek_qD,Bozek_2016_3q,Loizides_q,PHENIX1312.6676}.
This is due to the growth of the fraction of the wounded constituents in
each nucleon in $AA$ collisions as compared to that in $pp$ collisions.
In this picture the two component structure 
(\ref{eq:10}), supported by the data, 
 may simply be an empirical proxy for the pure quark-participant
scaling of the produced entropy without (or with very small) a binary collision
term at all \cite{PHENIX1312.6676}.
However, the recent analysis \cite{Bozek_2016_3q} 
shows that in the wounded quark model the contribution to multiplicity
from $qq$ interaction required for description of data on $AA$ collisions 
may differ substantially
from the one that is necessary for $pp$ collisions .
Say,  the data on Au+Au collisions at $\sqrt{s}=0.2$ TeV 
require the quark contribution suppressed by a factor of $\sim 1.4$ as 
compared to $pp$ interaction \cite{Bozek_2016_3q}. The results of 
\cite{Bozek_qD} show that the situation with consistency between 
$AA$ and $pp$ collisions becomes better if the nucleon is treated as a 
quark-diquark system.

In previous extensions of the standard wounded nucleon Glauber model to
the sub-nucleon level the sub-nucleon degrees of freedom have been 
assumed to be constituent quarks (or diquarks)
\cite{woun_q1,woun_q2,Voloshin_3q,Bozek_qD,Bozek_2016_3q,Loizides_q,PHENIX1312.6676}. 
However, it is well known that in the internal nucleon structure 
an important role is also played by the long-range meson-baryon fluctuations.
Analyses of the nucleon wave function in the infinite momentum 
frame (IMF) show that the total weight of the meson-baryon Fock states 
in the nucleon may be as large as $\sim 40$\% \cite{ST}.
These meson-baryon fluctuations are dominated by the 
$\pi N$  Fock component of the physical nucleon. 
It is known that the effect of the pion cloud plays an 
important role in diffractive processes
\cite{Drell,Deck,Kaidalov_dd}. In the presence of the meson-baryon
Fock components the diffraction excitation of the projectile proton 
emerges due to the well known Good-Walker mechanism
\cite{GW_dd} connected with the difference in elastic amplitudes for 
different Fock states. The meson-baryon Fock components are also important
for inclusive processes $pp\to n(\Delta^{++})X$ that are
dominated by inelastic interaction of the pion from the projectile
with the target proton \cite{Kaidalov_OPE1,Kaidalov_OPE2,Z_OPE1,Z_OPE2}.
It was understood long ago that 
the meson-baryon Fock components of the nucleon play an important 
role in the flavor dependence of nucleon parton distribution functions (PDFs) 
in deep inelastic scattering (DIS) \cite{ST}. It is believed  that the meson-baryon 
Fock components are responsible for the violation of the Gottfried 
sum rule \cite{ST}.
From the point of view of $AA$ collisions it is important
that, similarly to the wounded nucleon model
with constituent quarks, in the model with the meson degrees of freedom
there must be a nonlinear  increase of $dN_{ch}(AA)/d\eta$ 
with the number of the wounded nucleons.
It is clear that this effect should emerge independently of the
specific mechanism of inelastic processes.

In the present work we develop a MCG formalism which account for the
meson-baryon Fock component of the physical nucleon and   
address its possible effect on the entropy production in $AA$ collisions. We 
also study its effect in the formation of the small size plasma
fireball in $pA$ and $pp$ collisions. 
Following the studies on the effect of the meson cloud on the nucleon PDFs 
 \cite{ST} we use the IMF scheme for the meson-baryon
Fock states\footnote{Note that our calculations, from the point of view of the inelastic
cross sections correspond the account for the effect of the inelastic Gribov
rescatterings \cite{Gribov}. But one should bear in mind that from
the point of view of charged multiplicity density
the wounded nucleon scheme is not equivalent to calculations
in the Glauber model with the AGK rules that without Pomeron interactions give
for A+B nuclear collisions $\frac{dN_{ch}(AA)}{d\eta}=A B n_{pp}$ in the
central rapidity region \cite{Kaidalov_AGK}.}.
We will analyze within our MCG model 
the available data on the charged multiplicity in Au+Au 
collisions at $\sqrt{s}=0.2$ \cite{STAR1} and Pb+Pb collisions
at $2.76$ TeV \cite{ALICE1} and give predictions for Pb+Pb collisions at
$\sqrt{s}=5.02$ TeV for the future LHC run 2.

The plan of the paper is as follows. In Sec.~2 we 
discuss the IMF model for the physical nucleon.
In Sec.~2 we discuss the details of the MCG scheme.
In Sec.~4 we present our numerical results.
We first fix the 
parameters of the model from the charged multiplicity distribution in 
$pp$ collisions. Then we present the results of the MCG simulations 
of the charged multiplicity density and azimuthal asymmetry $\epsilon_2$ in 
$AA$ collisions. We also present the results for 
$pA$ and $pp$ collisions.
We give conclusions in Sec.~5.
Some of our results concerning the charged multiplicity
density for Au+Au and Pb+Pb collisions have been reported
in an earlier short communication \cite{Z_MCGL1}.

\section{Model for the meson-baryon component of the nucleon}
Our treatment of the meson-baryon component is similar to that used
in the analyses of the meson cloud effect on the nucleon PDFs
(for a nice review, see \cite{ST}) 
based on the IMF picture of the physical nucleon wave function.
At high energies this model is valid 
to the leading order in the nucleon energy.
We write the physical nucleon IMF wave function as the Fock state
composition of one- and two-body states 
\cite{ST,Zoller_MB}
\beq
|N_{phys}\rangle=\sqrt{1-n_{MB}}|N\rangle+
\sum_{MB}\int dxd\kb\Psi_{MB}(x,\kb)|MB\rangle\,.
\label{eq:40}
\eeq
Here $N$, $B$, and $M$ denote the bare baryon and meson states, 
$x$ is the fractional longitudinal meson momentum in the physical nucleon, 
$\kb$ is the tranverse meson momentum, $\Psi_{MB}$ is
the probability amplitude for the two-body $MB$ Fock state, and
\beq
n_{MB} =\sum_{MB}\int dxd\kb|\Psi_{MB}(x,\kb)|^2
\label{eq:50}
\eeq
is the total weight of the $MB$ Fock components. The dominant two-body Fock
component is the $\pi N$ state.
The IMF energy denominator of time-ordered perturbation theory 
for the $MB$ component  can be written as
$E_N-E_M-E_B\approx [m_N^2-M^{2}_{MB}(x,\kb^2)]/2E_N$, 
where
\beq
M^{2}_{MB}(x,\kb^2)=\frac{m_M^2+\kb^2}{x}+\frac{m^2_{B}+\kb^2}{1-x}
\label{eq:60}
\eeq
is the squared invariant mass of the two-body $MB$ system. 

The IMF wave function of the $MB$ Fock component 
(for point-like particles) may be written as 
\beq
\Psi_{MB}(x,\kb)=
\frac{\langle MB|V|N\rangle}
{4\pi^{3/2}\sqrt{x(1-x)}\left[m_N^2-M^2_{MB}(x,\kb)\right]}\,.
\label{eq:70}
\eeq
Eq. (\ref{eq:70}) corresponds to the wave function normalization 
of Eq. (\ref{eq:50}).
Here $\langle MB|V|N\rangle$ is the vertex factor in the IMF-limit, which
depends on the form of the Lagrangian. 
For the dominating $\pi N$ state the vertex reads 
$\langle \pi N'|V|N\rangle=g_{\pi NN}\bar{u}_{N'}\gamma_5u_{N}$  
(the helicity dependent vertex functions for different $MB$ states
can be found in \cite{ST}). 
The internal structure of the hadrons 
is accounted for by multiplying the vertex factor for point-like particles 
by a phenomenological formfactor, $F$.
To insure the charge and momentum conservation for the IMF wave functions
the formfactor should depend on $x$ and $\kb$ only via 
the invariant mass $M_{MB}(x,\kb)$ \cite{Zoller_MB1,Zoller_MB,ST,MST_MB}.
The information on the phenomenological formfactor $F$ for the dominating 
$\pi N$ component may be extracted from the data on the neutron 
spectrum in the process $pp\to nX$. 
The analysis \cite{HSS_MB} of the experimental neutron spectrum within
the IMF formalism with the dipole formfactor \cite{MST_MB} 
\beq
F=\left(\frac{\Lambda^2+m_N^2}{\Lambda^2+M^{2}_{\pi N}(x,\kb)}\right)^2\,
\label{eq:80}
\eeq
gives  $\Lambda\approx 1.3$ GeV. 
It is also supported by the data on the nucleon PDFs, because 
at the same time it allows to describe the violation of the Gottfried 
sum rule \cite{ST}. 
In Fig.~1 we show the $x$-distribution for the $\pi N$ state for this value
of the parameter $\Lambda$. 
One can see that the spectrum is peaked at $x\sim 0.3$.
For the $\rho$-meson the spectrum is peaked at somewhat larger 
$x$ ($x\sim 0.5$ \cite{ST}). 
\begin{figure}[ht]
\epsfig{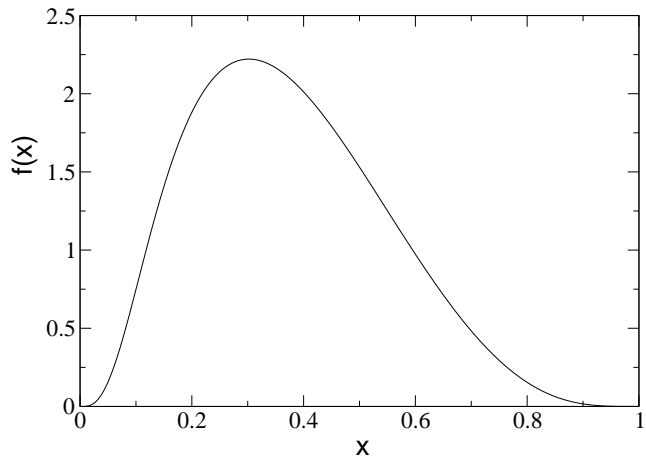}
\caption{\small 
Normalized to unity longitudinal $x$-distribution for 
the $\pi N$ Fock component obtained 
for the dipole formfactor (\ref{eq:80}) with $\Lambda=1.3$ GeV.
}
\end{figure}
\begin{figure}[ht]
\epsfig{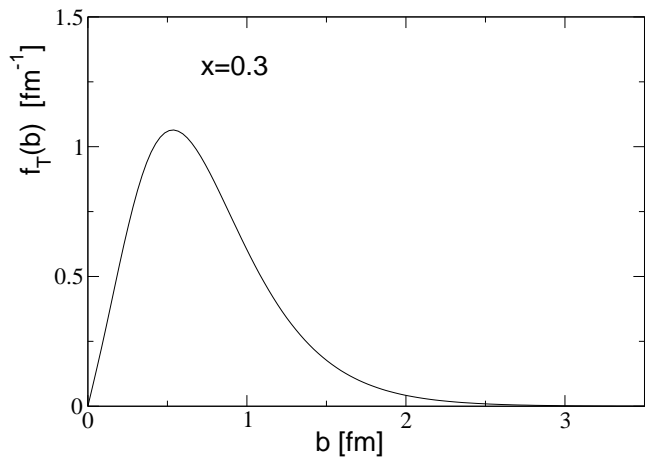}
\caption{\small 
Transverse distribution versus $b$ for the $\pi N$ state
at $x=0.3$ for the dipole formfactor (\ref{eq:80}) with $\Lambda=1.3$ GeV
with the normalization $\int db f_T(b)=1$.
}
\end{figure}
The transverse distribution for the $\pi N$ state at $x=0.3$ is shown in Fig.~2.
It gives for the root mean square transverse radius of the $\pi N$ component 
$\langle \rho^2_{\pi N}\rangle^{1/2}\approx0.87$ fm. 

The IMF scheme allows one to avoid the difficulties with
the momentum and charge conservation present in the earlier analyses 
of the effects of the meson cloud in DIS  
based on the covariant formulation \cite{Sul72,Speth_cov93}.
For the formfactor $F$ depending on the $M_{MB}(x,\kb)$
the meson and baryon $x$-distributions of the physical nucleon 
satisfy the relation
\beq
f_{M/N}(x)=f_{B/N}(1-x)\,,
\label{eq:90}
\eeq
where 
\beq
f_{M/N}(x)=\int d\kb |\Psi(x,\kb)|^2\,,
\label{eq:100}
\eeq
and  
\beq
f_{B/N}(x)=\int d\kb |\Psi(1-x,\kb)|^2\,.
\label{eq:110}
\eeq
The symmetry relation (\ref{eq:90}) between the $N\to M$ and 
$N\to B$ splitting functions is not satisfied in the covariant description
of the meson cloud with phenomenological formfactors depending on the 
invariant variable $t$ \cite{Zoller_MB1,ST}.

In DIS the meson and baryon
in the two-body Fock states act as an independent sources of the parton
distributions \cite{ST,MST_MB}. The results of the previous analyses of 
the meson cloud effects in DIS show that to a good accuracy in 
the Fock state decomposition (\ref{eq:20}) it is enough to include
$\pi N$, $\pi\Delta$, $\rho N$ and $\rho \Delta$ two-body systems.
The total weight of these 
four two-body states in the physical nucleon is
about $40$\% \cite{ST}. The representation (\ref{eq:40}) neglects the 
higher order terms from many-body systems. A qualitative 
analysis \cite{Zoller_MB1}
show that the effect of the higher order Fock states should be relatively
small for realistic formfactors.  
We assume that the soft inelastic $NN$ interaction may be described
as independent inelastic interactions of the bare meson and baryon constituents
of the colliding nucleons. The specific mechanism of the inelastic
interactions of the bare constituents is not crucial at 
this point\footnote{We restrict ourselves to the leading order
contribution from the meson and baryon degrees of freedom of the physical 
nucleon in the IMF. 
Of course, each meson/baryon constituent develops its own 
IMF wave function at the quark-gluon level, that is important from
the point of view the inelastic interactions of the meson/baryon 
constituents. As it occurs, say, in the IT-Glasma model 
\cite{IP-GL1,IP-GL2}. In principle, higher order meson/baryon Fock states
in the IMF wave function also can contribute to the inelastic interactions 
of the bare meson/baryon states. However, in the wounded nucleon Glauber
model this complicated dynamics is replaced by the simple 
phenomenological ansatz on the entropy production from inelastic interactions
of the bare meson/baryon constituents. 
}.  
Because the quark contents of the bare $\Delta$ and $\rho$-meson are
the same as for the bare $N$ and $\pi$ states, we
make a reasonable assumption that from the point of view of 
inelastic interactions the bare $\Delta$ is equivalent to the 
bare $N$ state and the $\rho$-meson is equivalent the pion.
Then, in the wounded nucleon picture 
each physical nucleon interacts with the probability $1-n_{MB}$ as the bare $N$ and with the probability $n_{MB}$ as the two-body $\pi N$ system.
Since the dominating $\pi N$ state is peaked at $x\approx 0.3$
we, for the sake of simplicity, take for the fractional meson momentum 
in the effective $MB$ component 
$x=0.3$. 
For the transverse spacial distribution of the $MB$ state
we use the distribution of the dominant $\pi N$ component at $x=0.3$
shown in Fig. 2.
We renormalized it to match the total weight of the 
$\pi N$, $\pi\Delta$, $\rho N$ and $\rho \Delta$ components
$n_{MB}=0.4$ \cite{ST}.  
Note that the results 
of the MCG simulation are not very sensitive to the value of $\Lambda$.
This is because there is no shadowing effects for inelastic interactions
of the baryon and meson constituents.

\section{Formulation of the MCG scheme}
For the two component model of the nucleon (\ref{eq:40}) 
inelastic interaction of the physical nucleons
from the colliding objects occurs as $N+N$, $N+MB$, $MB+N$ and
$MB+MB$ collisions. We assume that the inelastic cross sections
for the bare states obey the constituent quark counting rule
$4\sigma^{NN}_{in}=6\sigma^{MB}_{in}=9\sigma^{MM}_{in}$.
For the impact parameter  profile of the probability of $ab$ 
inelastic interaction we use a Gaussian form 
\beq
P_{ab}(\rho)=\exp\left(-\pi \rho^2/\sigma_{in}^{ab}\right)\,.
\label{eq:120}
\eeq
We adjusted the value of the parameter $\sigma_{in}^{NN}$ to reproduce 
the experimental inelastic $pp$ cross section $\sigma_{in}^{pp}$ (see below).

We consider the charged multiplicity density $dN_{ch}/d\eta$
at the central pseudorapidity $\eta=0$
(sometimes, for clarity, we will use for $dN_{ch}/d\eta$ a 
simple notation $N_{ch}$ assuming that 
the charged multiplicity is defined in the unit pseudorapidity 
window $|\eta|<0.5$).
For calculation of 
the contribution to the multiplicity density from the $MB$ component
we need to know the $dN_{ch}/d\eta$ for pion-nucleon and pion-pion
collisions. The direct data for pion-proton and pion-pion 
collisions for RHIC-LHC energies are absent. 
We use the information from the quark-gluon string
model \cite{Kaidalov,Capella}. Calculations within this model show 
that the charged particle multiplicity density in the central rapidity region for $\pi p$ and
$\pi\pi$ collisions is somewhat bigger than that in $pp$ collisions. 
Our calculations show that to good accuracy this small excess 
compensates a possible reduction of the multiplicity
density in $\pi p$ and $\pi \pi$ interactions due to somewhat smaller 
c.m. energy in our model.
For this reason we, for the sake of simplicity, assume that all the 
wounded bare particles produce the same amount of entropy per unit 
rapidity in the c.m. frame of colliding objects.
We ignore the effect of a small rapidity shift ($\sim 0.5$) of the c.m. frame 
for pairs with different energies (as it occurs for $\pi N$ interactions)
on the entropy rapidity density, because the charged multiplicity density
is almost flat at midrapidity.

The total entropy rapidity density of the fireball for the A+B collision
is the sum of the contributions from the sources corresponding to 
the wounded constituents and to the binary collisions of the constituents 
\beq
\frac{dS}{dy}=\sum_{i=1}^{N_w} \frac{dS_w^{i}}{dy}+
\sum_{i=1}^{N_{bin}} \frac{dS_{bin}^{i}}{dy}\,,
\label{eq:130}
\eeq
where 
\beq
\frac{dS_w^i}{dy}=\frac{(1-\alpha)}{2}S
\label{eq:140}
\eeq
is the contribution of individual source from the 
wounded constituents in the systems A and B, and
\beq
\frac{dS_{bin}^i}{dy}=S
\label{eq:150}
\eeq
is the contribution of individual binary collision.
In the MCG simulations we assume that for each pair of 
wounded particles the probability of a hard binary collision is $\alpha$.
As usually done in the MCG schemes, to model the fluctuations of 
the multiplicity density in $pp$ collisions,
we treat the quantity $S$ in (\ref{eq:140}), (\ref{eq:150}), as a random variable.
We assume an isentropic evolution of the fireball. 
For the isentropic expansion the initial entropy rapidity density 
is proportional to the final charged
multiplicity pseudorapidity density
\beq
dS/dy=C dN_{ch}/d\eta\,,
\label{eq:160}
\eeq
where $C\approx 7.67$ \cite{BM-entropy}. 
In this approximation
we can work directly with the pseudorapidity charged particle density
$dN_{ch}/d\eta$.
So we will treat each fluctuating entropy source as 
a source producing a fluctuating amount $n=S/C$ of charged particles
in the unit pseudorapidity  interval $|\eta|<0.5$. 
We describe the fluctuations of $n$ for each source by the Gamma distribution
\beq
\Gamma(n,\langle n\rangle)=
\left(\frac{n}{\langle n\rangle}\right)^{\kappa-1}
\frac{\kappa^\kappa\exp\left[-n\kappa/\langle n\rangle\right]}
{\langle n\rangle \Gamma(\kappa)}\,,
\label{eq:170}
\eeq
which is widely used in the MCG simulations.
We adjusted the parameters $\langle n\rangle$ and $\kappa$ to
the experimental $pp$ data on the mean charged multiplicity 
and its variance in the unit pseudorapidity window $|\eta|<0.5$
(see below).
Our calculations show that for $AA$ 
collisions the results for the Gamma distributions are very similar 
to that for the negative binominal distribution. 

In comparison with experimental data we, as usual, define the 
centrality $c$ in $AA$ collisions through the theoretical charged multiplicity 
distribution $P$ \cite{Broniowski_c}
\beq
c(N_{ch})=\sum_{N=N_{ch}}^{\infty}P(N)\,.
\label{eq:180}
\eeq
Here $N_{ch}$ is the theoretical charged multiplicity for $|\eta|<0.5$,
i.e. $dN_{ch}/d\eta$ in our MCG simulations.
For calculation of the centrality dependence of the charged particle
multiplicity the distribution of the entropy
rapidity density in the transverse coordinates
is not important. 
However, the spacial distribution of the deposited entropy 
is important for the geometric quantities, such as the initial 
anisotropy coefficients $\epsilon_n$ of the fireball. 
In terms of the spacial entropy distribution
(we denote it by $\rho_s=dS/dyd\ro$) the coefficients $\epsilon_n$ read
\cite{Teaney_en,Ollitraut_en}
\beq
\epsilon_n=\frac{\left|\int d\ro \rho^n e^{in\phi}\rho_s(\ro)\right|}
{\int d\ro \rho^n\rho_s(\ro)}\,
\label{eq:190}
\eeq
(here it is assumed that the transverse vectors $\ro$ are calculated in the 
tranverse c.m. frame, i.e., $\int d\ro \ro\rho_s(\ro)=0$). 
For the point-like sources we have
\beq
\rho_s(\ro)
=\sum_{i=1}^{N_w} \delta(\ro-\ro_i)\frac{dS_w^{i}}{dy}+
\sum_{i=1}^{N_{bin}} \delta(\ro-\ro_i)\frac{dS_{bin}^{i}}{dy}\,.
\label{eq:200}
\eeq
We use the popular prescription that 
for the wounded constituents the centers of the sources are located at the 
positions of the wounded particles, and that for 
each binary collision the source is located in the middle between 
colliding particles. 
Of course, physically, the approximation of the point-like sources is
clearly unreasonable. We account for qualitatively the finite size
of the sources by replacing 
the $\delta$ functions
in (\ref{eq:200}) by a Gaussian distribution
\beq  
\exp{\left(-\ro^2/\sigma^2\right)}/\pi \sigma^2\,.
\label{eq:210}
\eeq
We perform calculations for
$\sigma=0.7$ and $0.4$ fm.
The results for the anisotropy coefficients in $AA$ collisions 
become sensitive 
to the width of smearing of the sources only for very peripheral 
collisions, but for the small size fireballs in $pA$ and $pp$ 
collisions the value of $\sigma$ is very crucial (see below).

We perform calculations for the Woods-Saxon nuclear distribution.
We account for the deformation for the nuclei $^{197}$Au and $^{238}$U.
For these nuclei we use the $\theta$-dependent Woods-Saxon nuclear density
\beq
\rho_{A}(r,\theta)=\frac{\rho_0}{1+\exp[(r-R_A(\theta)/a]}\,,
\label{eq:220}
\eeq
\beq
R_A(\theta)=R[1+\beta_2Y_{20}(\theta)+\beta_4Y_{40}(\theta)]
\label{eq:230}
\eeq
with $Y_{20}$ and $Y_{40}$ the spherical harmonics. 
Following \cite{GLISS2}  we take $R=6.37(6.8)$ fm, $\beta_2=-0.13(0.28)$, and 
$\beta_4=-0.03(0.093)$  for Au(U) nuclei, and $a=0.54$ fm.
For the $^{207}$Pb nucleus we use the ordinary spherically symmetrical
Woods-Saxon formula
with 
a $\theta$-independent ($\beta_{02}=\beta_{04}=0$) radius
$R_{A}=(1.12A^{1/3}-0.86/A^{1/3})=6.49$ fm, and $a=0.54$ fm \cite{GLISS2}.

\section{Numerical results}
\subsection{Parameters of the model for $pp$ collisions}
In numerical calculations for $pp$ collisions
we take $n_{pp}=2.65$ at $\sqrt{s}=0.2$ TeV
obtained by the UA1 collaboration \cite{UA1_pp} for non-single-diffractive
(NSD) events. 
\begin{figure}[ht]
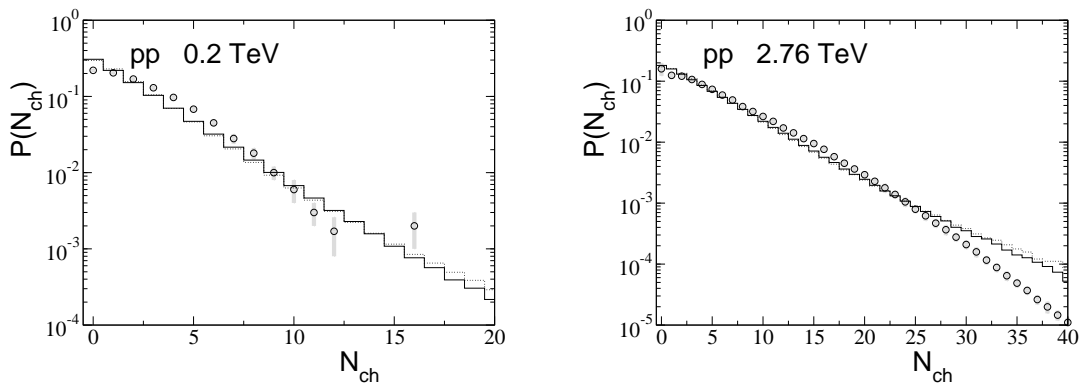

\hspace*{-0.8cm }\epsfig{file=fig3a.eps,height=5cm,clip=,angle=0} 
\hspace*{0.8cm } \epsfig{file=fig3b.eps,height=5cm,clip=,angle=0} 
\caption{\small 
Charged multiplicity distribution in
$pp$ collisions 
for the pseudorapidity window $|\eta|<0.5$. 
Left: MCG simulation for $\sqrt{s}=0.2$ TeV  
for the scenario with the meson cloud for $\alpha=0.06$ (solid) 
and without the meson cloud
for $\alpha=0.135$ (dotted), the data are from UA5 
\cite{UA5_ZP1989}.
Right: MCG simulation for $\sqrt{s}=2.76$ TeV  for the scenario with the meson cloud for $\alpha=0.09$ (solid) 
and without the meson cloud
for $\alpha=0.14$ (dotted), the data are from ALICE 
\cite{ALICE_nch541}.
}
\end{figure}
For NSD $pp$ events at $\sqrt{s}=2.76$ TeV we use the ALICE 
result $n_{pp}\approx 4.63$ \cite{ALICE_nch541} .
In the MCG simulations 
for $\sigma_{in}^{pp}$ 
we also use the inelastic $pp$ cross section corresponding to the NSD 
event class.
The exclusion of the diffractive contribution 
is reasonable because the diffractive events do not contribute to the 
midrapidity multiplicity density considered in the present work.
We use for the NSD $pp$ inelastic cross section
at $\sqrt{s}=0.2$ TeV
the value $35$ mb measured by the UA1 collaboration \cite{UA1_pp}.
For $\sqrt{s}=2.76$ TeV we use
the value $50.24$ mb obtained by the ALICE collaboration
\cite{ALICE4968}.
With the above values of the NSD $\sigma_{in}^{pp}$
we fitted  $\sigma_{in}^{NN}$ for the bare nucleons  necessary for 
the MCG simulations 
in the version with the $MB$ Fock component. 
For this version we obtained 
\beq
\sigma_{in}^{NN}[\sqrt{s}=0.2\,,2.76\,\mbox{TeV}] \approx [26.15,\,38.4]
\,\,\mbox{mb}\,.
\label{eq:240}
\eeq
For the version without the meson cloud the parameter $\sigma_{in}^{NN}$ 
is equal 
simply to  the experimental NSD $pp$ cross section.
We fitted the parameters $\langle n\rangle$ and $\kappa$ of the 
Gamma distribution (\ref{eq:170}) to reproduce the experimental 
mean $N_{ch}$ in $pp$ collisions and to satisfy the relation $N_{ch}/D=1$ 
($D^2$ is a variance of
$N_{ch}$) in the pseudorapidity window $|\eta|<0.5$,  which 
is well satisfied for the experimental multiplicity distribution
both at $\sqrt{s}=0.2$ TeV \cite{UA1_pp} and at 
$\sqrt{s}=2.76$ TeV \cite{ALICE_nch541}.
For the scenario without the meson cloud the parameter $\langle n\rangle$
should be equal to the experimental $dN_{ch}/d\eta$ for any 
fraction of the binary collisions $\alpha$. But the value of the parameter 
$\kappa$ depends on $\alpha$. At $\alpha=0$ the relation $N_{ch}/D=1$ gives 
$\kappa=0.5$.
For $\alpha>0$ the value of $\kappa$ grows weakly with $\alpha$. 
But the deviation from $0.5$ is relatively small. For the scenario with
the $MB$ component $\kappa$ is also close to $0.5$, and the required value 
of $\langle n\rangle$ is smaller than the experimental mean $N_{ch}$.  

To determine the values of the parameter $\alpha$ we used a two step procedure.
First, we fitted the parameters $\langle n\rangle$ and $\kappa$ to 
the $pp$ data on $N_{ch}$  imposing the condition $N_{ch}/D=1$ 
for a broad set of $\alpha$ from $0$ to $0.2$. 
In the second step, we used the set of $\langle n\rangle$ and $\kappa$ to fit 
the parameter $\alpha$ to best reproduce the
data on the centrality dependence of the midrapidity $dN_{ch}/d\eta$
in Au+Au collisions at $\sqrt{s}=0.2$ TeV from STAR \cite{STAR1}
and in Pb+Pb collisions at $\sqrt{s}=2.76$ TeV from ALICE \cite{ALICE1}.
For Au+Au collisions at $\sqrt{s}=0.2$ TeV this procedure gives 
$\alpha\approx0.06$ and $\alpha\approx 0.135$  for the scenarios 
with and without the meson cloud, respectively.
From the ALICE data on Pb+Pb collisions at $\sqrt{s}=2.76$ TeV \cite{ALICE1}
we obtained $\alpha\approx0.09$ and $\alpha\approx 0.14$ for the scenarios
with and without the meson cloud, respectively.
The parameters of the Gamma distribution
(\ref{eq:170}) obtained from the fit with the meson 
cloud to the $pp$ data for the above optimal values of $\alpha$ 
read
\beq
\langle n\rangle[\sqrt{s}=0.2,\,2.76\,\text{TeV}]\approx
[2.39,\,4.13]\,,
\label{eq:250}
\eeq
\beq
\kappa[\sqrt{s}=0.2,\,2.76\,\text{TeV}]\approx
[0.506,\,0.52]\,.
\label{eq:260}
\eeq
For the scenario without the meson cloud for the optimal values
of $\alpha$ we obtained
\beq
\kappa[\sqrt{s}=0.2,\,2.76\,\text{TeV}]\approx
[0.57,\,0.57]\,,
\label{eq:270}
\eeq
(as we said $\langle n\rangle$ is equal to the experimental $n_{pp}$).
As one could expect a priori, 
accounting for the meson cloud leads to a reduction of the
required fraction of the binary collisions. The effect becomes 
somewhat smaller at the LHC energy. This is natural because the 
interaction radius becomes bigger at the LHC energies. It results in 
a lower sensitivity to the internal nucleon structure.

In Fig.~3 we compare the multiplicity distribution of charged hadrons
for the versions with and without the meson cloud
for $|\eta|<0.5$ for $pp$ collisions at $\sqrt{s}=0.2$ TeV and $\sqrt{s}=2.76$
TeV with the experimental data from UA5 
\cite{UA5_ZP1989} and ALICE \cite{ALICE_nch541}.
One sees that for both of the versions the agreement with the data are 
reasonable in the region $N_{ch}\lsim 5\langle N_{ch}\rangle$. 
Note however, that the multiplicity 
distributions in $AA$ collisions is not very sensitive to the specific form 
of the multiplicity distribution in $pp$ collisions except for the 
region of very peripheral collisions when the number of the wounded nucleons
becomes small. Anyway, the tail region of the charged multiplicity 
distribution with $N_{ch}\gg\langle N_{ch}\rangle$ practically cannot
affect the theoretical predictions for $AA$ collisions.

\subsection{A+A collisions: charged multiplicity density}
In Figs.~4,~5 we compare our results for centrality
dependence of the charged multiplicity density $dN_{ch}/d\eta$ at $\eta=0$ 
for the fitted values of $\alpha$ with the STAR 
data on Au+Au collisions at $\sqrt{s}=0.2$ TeV \cite{STAR1} and the 
ALICE data on Pb+Pb collisions at $\sqrt{s}=2.76$ TeV \cite{ALICE1}. 
The theoretical histograms have been obtained by Monte Carlo generation 
of samples with $\sim 2\times 10^6$ events. As was noted earlier, 
our parameters of the Gamma distribution (\ref{eq:170}) and the definition
of centrality for $AA$ collisions (\ref{eq:180}) correspond to
the unit pseudorapidity window $|\eta|<0.5$. The centrality categorization 
in the STAR data \cite{STAR1} 
and the ALICE data \cite{ALICE1}
is also performed via the charged multiplicity at $|\eta|<0.5$. 
However, in principle, 
for collisions of heavy nuclei it is not very crucial because
the effect of the multiplicity fluctuations at a given impact parameter
(except for very peripheral collisions) on the centrality categorization of 
a given event is small \cite{Broniowski_c}. 
To illustrate better the magnitude of the effect of the meson cloud
we show in Figs.~4a and 5a the results for the scenario without the meson 
cloud, but obtained with the optimal $\alpha$ for the scenario 
with the meson cloud.
Comparison of the two histograms show that at small centrality 
the meson cloud increases the multiplicity by $\sim 16-18$\%.
Since our calculations do not assume a certain mechanism of the entropy 
production in collisions of the bare baryon and meson states, one 
can expect that the $MB$ Fock components in the nucleon wave function
should increase the multiplicity in $AA$ collisions 
in any scheme. 
\begin{figure}[ht]
\epsfig{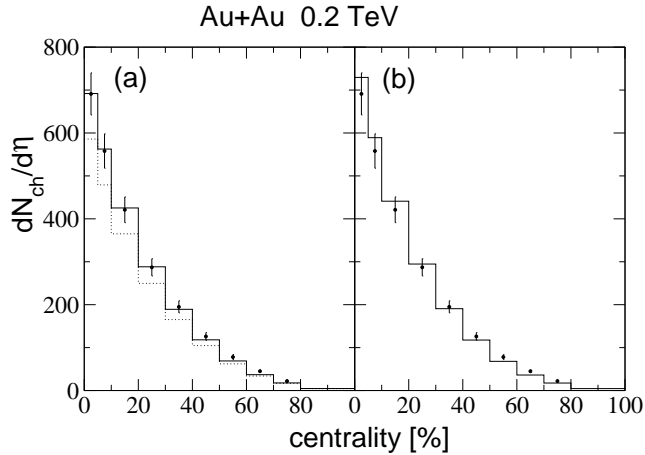}
\caption{\small Centrality dependence of $dN_{ch}/d\eta$ at $\eta=0$ for 
Au+Au collisions at $\sqrt{s}=0.2$ TeV. Left: 
MCG simulation for the scenarios with (solid) and 
without (dotted) the meson cloud
for $\alpha=0.06$.
Right: MCG simulation for the scenario without the meson cloud
for $\alpha=0.135$. Data are from STAR \cite{STAR1}.
}
\end{figure}
\begin{figure}%[ht]
\epsfig{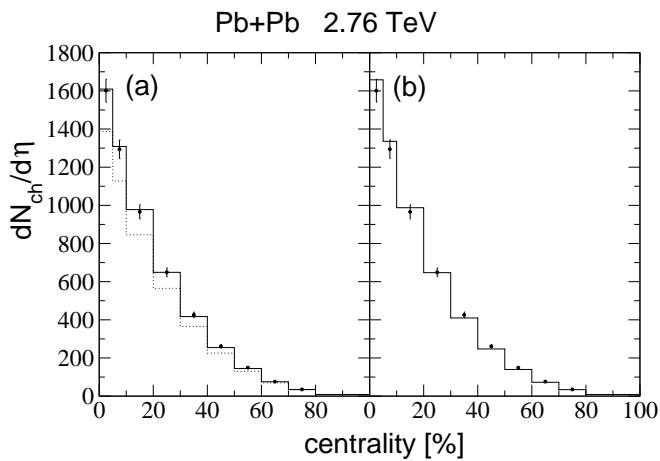}
\caption{\small Centrality dependence of $dN_{ch}/d\eta$ for 
Pb+Pb collisions at $\sqrt{s}=2.76$ TeV. Left: 
MCG simulation for the scenarios with (solid) and 
without (dotted) the meson cloud
for $\alpha=0.09$.
Right: MCG simulation for the scenario without the meson cloud
for $\alpha=0.14$. Data are from ALICE \cite{ALICE1}.
}
\end{figure}

In the future LHC run 2  Pb+Pb collisions will be studied at
$\sqrt{s}=5.02$ TeV. To give the theoretical prediction
for Pb+Pb collisions at this energy 
we have used the same values of the parameter $\alpha$ as 
for $\sqrt{s}=2.76$ TeV. Since the
variation of $\alpha$ from $\sqrt{s}=0.2$ TeV to $\sqrt{s}=2.76$ TeV
is not strong, one can expect that its variation from  $\sqrt{s}=2.76$ TeV to 
$\sqrt{s}=5.02$ TeV should not be significant.
The direct $pp$ data on $dN_{ch}/d\eta$ at 
$\sqrt{s}=5.02$ TeV are absent. 
We obtained it with the help of the power law interpolation 
$dN_{ch}/d\eta \propto s^{\delta}$
between the ALICE data \cite{ALICE_nch541} at $\sqrt{s}=2.76$ TeV
($dN_{ch}/d\eta \approx 4.63$)  and 
at $\sqrt{s}=7$ TeV   
($dN_{ch}/d\eta=5.74\pm 0.15$). 
It gives $dN_{ch}/d\eta\approx 5.35$ at $\sqrt{s}=5.02$ TeV.
We use for the NSD $pp$ inelastic cross section
at $\sqrt{s}=5.02$ TeV the value $55.44$ mb obtained
by interpolating between the ALICE data \cite{ALICE4968}  at 
$\sqrt{s}=2.76$ and $7$ TeV.
Making use of the above NSD $\sigma_{in}^{pp}$
we fitted the parameter $\sigma_{in}^{NN}$ for the scenario with the meson 
cloud for $\alpha=0.09$. We obtained 
$\sigma_{in}^{NN} (\sqrt{s}=5.02\,\mbox{TeV}) \approx 42.49$ mb
(as in the analysis of for $\sqrt{s}=2.76$ TeV 
for the scenario without the meson cloud $\sigma_{in}^{NN}$ is equal 
to  the NSD $pp$ cross section).
As above, the parameters $\langle n\rangle$ and $\kappa$ of the Gamma distribution 
(\ref{eq:170}) have 
been fitted to reproduce the experimental $n_{pp}$ and to satisfy the
relation $n_{pp}/D=1$. Without the meson cloud $\langle n\rangle$
is simply equal to 
the interpolation of the experimental $dN_{ch}/d\eta$ 
between $\sqrt{s}=2.76$ and $7$ TeV,
and fit of $\kappa$ for $\alpha=0.14$ gives $\kappa=0.564$.
For the scenario with the meson cloud we obtained 
$\langle n\rangle\approx4.72$, and $\kappa\approx 0.52$.
In Fig. 6 we compare our results for centrality dependence of the
charged multiplicity in Pb+Pb 
at $\sqrt{s}=5.02$ TeV
and at $\sqrt{s}=2.76$ TeV.
From Fig. 6 one sees that as compared to $\sqrt{s}=2.76$  
the growth of $dN_{ch}/d\eta$
in the central Pb+Pb collisions for $\sqrt{s}=5.02$ TeV is 
about $20$\%. It corresponds to increase of the fireball
initial temperature by $\sim 6$\%. 
\begin{figure}%[ht]
\epsfig{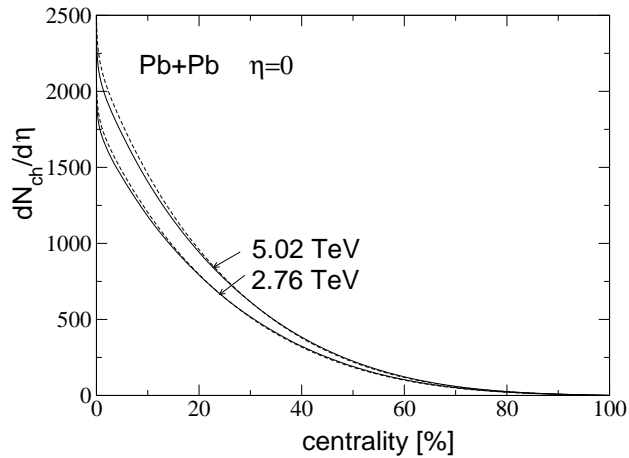}
\caption{\small Comparison of the centrality dependence of 
$dN_{ch}/d\eta$ at $\eta=0$ for Pb+Pb collisions at $\sqrt{s}=2.76$ 
and $5.02$ TeV
obtained from the MCG simulation 
for the scenarios with (solid) the meson cloud for $\alpha=0.09$ 
and without (dashed) the meson cloud for $\alpha=0.14$.
}
\end{figure}
\begin{figure}[h]
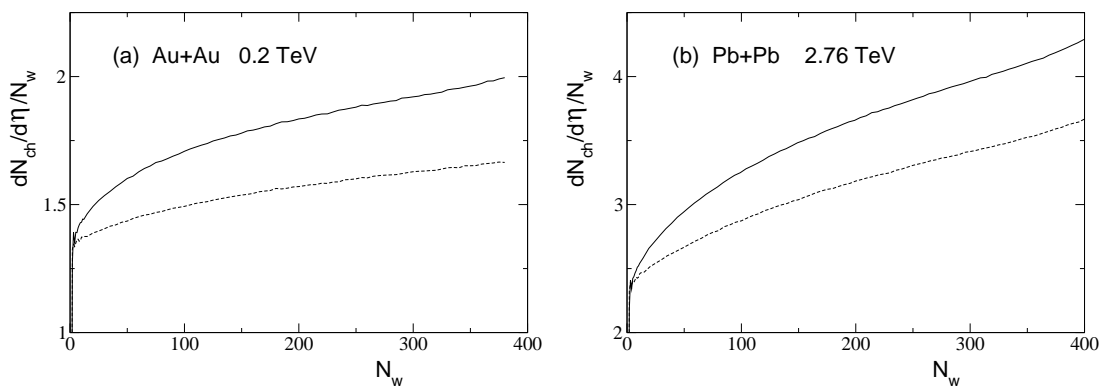

\epsfig{file=fig7a.eps,height=5.cm,clip=}\hspace{.2cm}
\epsfig{file=fig7b.eps,height=5cm,clip=}
\caption{\small Ratio $dN_{ch}/d\eta/N_w$ versus the number of the wounded
  nucleons $N_w$. Left: 
MCG simulation 
for Au+Au collisions at $\sqrt{s}=0.2$ TeV
for the scenarios with (solid) and without (dashed) the meson cloud
for $\alpha=0.06$.
Right: MCG simulation 
for Pb+Pb collisions at $\sqrt{s}=2.76$ TeV
for the scenarios 
with (solid) and without (dashed) the meson cloud
for $\alpha=0.09$. 
}
\end{figure}
\begin{figure}[ht]
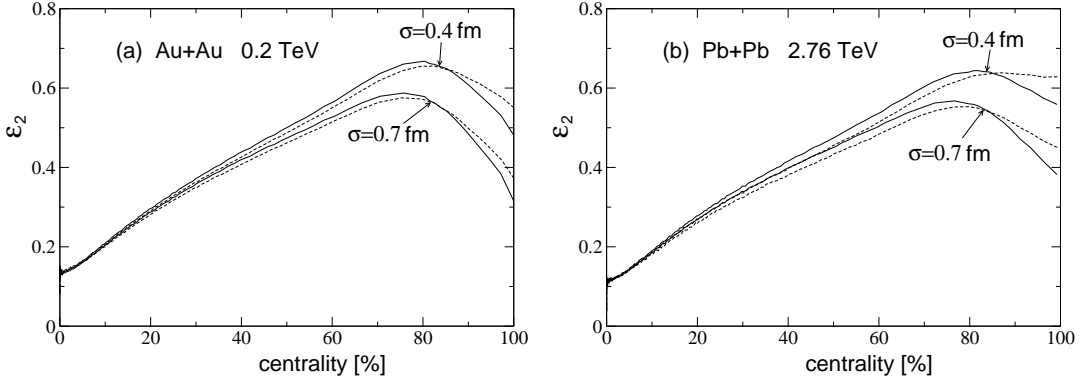

\epsfig{file=fig8a.eps,height=5.cm,clip=}\hspace{.2cm}
\epsfig{file=fig8b.eps,height=5cm,clip=}
\caption{\small Centrality dependence of the rms $\epsilon_2$ for
the Gaussian source distribution (\ref{eq:210}) for $\sigma=0.7$
and $0.4$ fm. 
Left: MCG simulation 
for Au+Au collisions at $\sqrt{s}=0.2$ TeV
for the scenarios with (solid) the meson cloud 
for $\alpha=0.06$
and without (dashed) the meson cloud for $\alpha=0.135$.
Right: MCG simulation 
for Pb+Pb collisions at $\sqrt{s}=2.76$ TeV
for the scenarios with (solid) the meson cloud 
for $\alpha=0.09$
and without (dashed) the meson cloud for $\alpha=0.14$.
}
\end{figure}
\begin{figure}
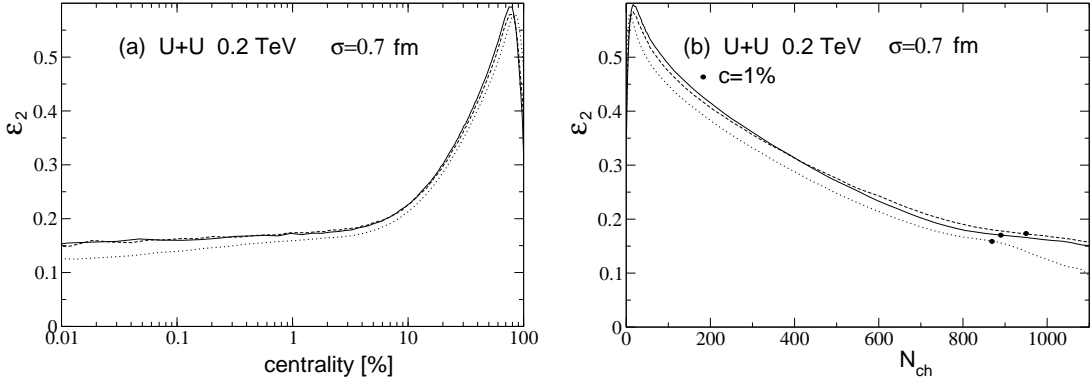
%[ht]
\epsfig{file=fig9a.eps,height=5cm,clip=}
\hspace{.2cm}
\epsfig{file=fig9b.eps,height=5cm,clip=}
\caption{\small Centrality (left) and $N_{ch}$ (right) 
dependence of the rms $\epsilon_2$ for 
U+U collisions at $\sqrt{s}=0.2$ TeV obtained for the Gaussian 
source distribution (\ref{eq:210}) for $\sigma=0.7$ fm. 
Solid: MCG simulation for the scenario with the meson cloud
for $\alpha=0.06$.
Dashed: MCG simulation for the scenario without the meson cloud
for $\alpha=0.135$.
Dotted: MCG simulation for the scenario without the meson cloud
for $\alpha=0.135$ without fluctuations of the charged multiplicity 
in $NN$ collisions for $n_{pp}=2.65$. In the right panel the dots 
mark the points with $c=1$\%.
}
\end{figure}
\begin{figure}[ht]
\epsfig{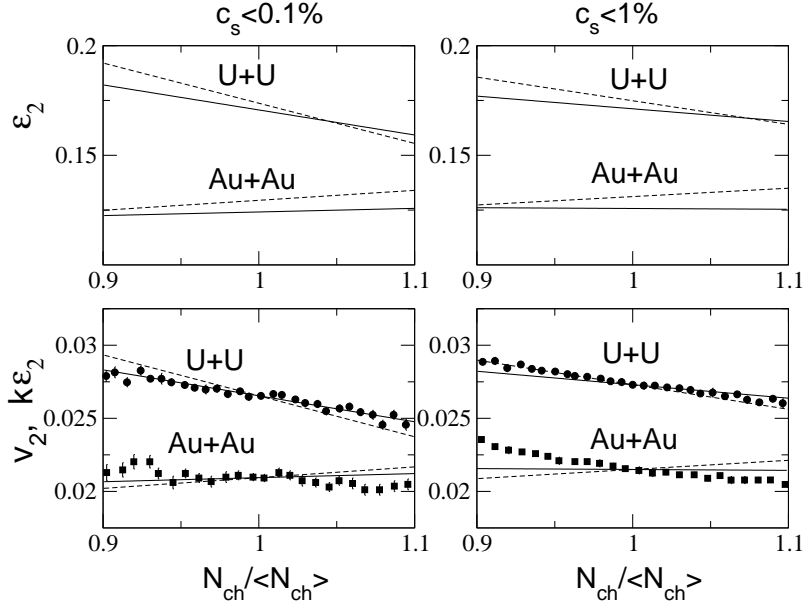}
\caption{\small rms ellipticity $\epsilon_2$ (upper panels)
and $v_2$ (lower panels) versus $N_{ch}/\langle N_{ch}\rangle$
in Au+Au and U+U collisions  
for the spectator centrality $c_s<0.1$\% (left) and 
$c_s<1$\% (right). 
In the upper panels the solid lines are the linear fits of our 
MCG simulation for $\sigma=0.7$ fm  with the $MB$ component for $\alpha=0.06$,
and the dashed lines are the same but without the $MB$ component
for $\alpha=0.135$. 
The lower panels show the comparison 
of our results with the STAR data \cite{UU_STAR} on $v_2$ 
in Au+Au collisions
at $\sqrt{s}=0.2$ TeV and U+U collisions at $\sqrt{s}=0.193$ TeV.
The solid and dashed lines show the MCG results for $k\epsilon_2$ with 
the renormalization constant $k$ defined to match the experimental $v_2$ at $N_{ch}=\langle N_{ch}\rangle$.
The lines have the same meaning as in the upper
panels.
}
\end{figure}

As was noted, in the scheme with the meson cloud the contribution of 
soft processes 
to the multiplicity density is not proportional 
to the number of wounded nucleons, because the probability 
of inelastic interactions of the meson states depends on centrality.
To illustrate the variation with centrality of the contribution of each
wounded nucleon due to the $MB$ component in Fig.~7 we plot the 
ratio $dN_{ch}/d\eta/N_w$ as a function of $N_w$
obtained with and without the $MB$ component for the  same $\alpha$
($\alpha=0.06$ for Au+Au collisions at $\sqrt{s}=0.2$ TeV and
$\alpha=0.09$ for Pb+Pb collisions at $\sqrt{s}=2.76$ TeV).
One sees that from peripheral to central
collisions this ratio rises by $\sim 20$\% and $\sim 15$\% for RHIC
and LHC conditions, respectively.

\subsection{A+A collisions: azimuthal eccentricity of the fireball}
We have also studied the effect of the $MB$ component on 
the root mean square (rms) anisotropy coefficient $\epsilon_2$
(which is often noted as 
$
\epsilon_{2}\{2\} %=\sqrt{\langle \epsilon_2^2\rangle}
$
).
In Fig. 8 we present the results for 
the rms $\epsilon_2$ versus centrality for
Au+Au at $\sqrt{s}=0.2$ TeV and Pb+Pb at $\sqrt{s}=2.76$ TeV
for the two models. 
We present the results for two values of  
the Gaussian width of the sources $\sigma=0.7$ and $0.4$ fm.
One sees that for small centrality the results with and without
the meson cloud are close to each other. For noncentral
collisions with centrality $\lsim 80$\% the version with the meson cloud
gives a little smaller $\epsilon_2$. But for very peripheral collisions
with centrality $\gsim 80$\% the anisotropy for the model with the meson
cloud becomes bigger than that without the meson cloud. The dependence
of the asymmetry on $\sigma$  is rather weak, except for the region
of large centrality ($\gsim 60-70$\%), where the typical number of 
sources is small and the results become sensitive to the shape 
of the entropy distribution in individual $NN$ collisions.  
Of course, in this region the results are very model dependent 
and not robust.
Thus we found that the effect of the meson cloud on 
the eccentricity $\epsilon_2$ in $AA$ collisions is relatively small.

Recently there has been considerable interest in the centrality/multiplicity 
dependence of the eccentricity $\epsilon_2$ for U+U collisions 
\cite{UU_Lednicky,UU_Voloshin,UU_Bass,UU_Singh} due to expected sensitivity
of the multiplicity to orientation of the colliding nuclei
connected with the prolate shape of the $^{238}$U-nucleus.
In \cite{UU_Lednicky,UU_Voloshin}
it was predicted that in U+U collisions the initial asymmetry 
$\epsilon_2$ should have a knee structure at multiplicities
in the top 1\% central U+U collisions (the knee-like structure has also 
been found in the recent MCG simulation of \cite{UU_Singh}). It may be interpreted 
as due to the 
growth of the relative contribution of the binary collisions for 
the tip-tip configurations of the colliding nuclei for a nonzero fraction 
$\alpha$ of the $\propto N_{coll}$ term in the MCG scheme. 
However the elliptic flow $v_2$ measured by STAR \cite{UU_STAR} in
U+U collisions at $\sqrt{s}=193$ GeV shows no indication of such a 
knee structure. This challenged the two component Glauber model with a 
significant contribution of the binary collisions, and stimulated searches 
for  alternative ansatze for the entropy generation in the Glauber picture
\cite{UU_Bass,UU_Singh}.
But it worth noting that the theoretical situation with the 
knee in the $\epsilon_2$ for the standard two component MCG model  
is still somewhat controversial. 
Indeed, the analysis \cite{UU_Lednicky} has been performed neglecting the 
fluctuations of the multiplicity in $NN$ collisions.
But in \cite{UU_Voloshin,UU_Singh} the fluctuations of the charged multiplicity
in $NN$ collisions have been taken into account.  
However, later in \cite{UU_Broniowski} it was demonstrated that
the knee structure vanishes when the fluctuations are taken into account,
that is in contradiction with the analyses \cite{UU_Voloshin,UU_Singh}.
On the other hand, more recent analysis \cite{UU_Heinz} indicates on 
appearance of a weak knee-like
feature even when the multiplicity fluctuations in $NN$ collisions
are taken into account.  
In Fig.~9a we show our results for $\epsilon_2$ versus centrality 
in U+U collisions at 
$\sqrt{s}=0.2$ TeV for the source width $\sigma=0.7$ fm. To stretch
the region of small centralities a logarithmic scale is used.
One can see that the predictions for 
the rms $\epsilon_2$ in the centrality region $0.01\lsim c\lsim 1$\% 
for the versions with and without the $MB$ component are very similar.
In this region $\epsilon_2$ decreases very smoothly with $c$, and 
does not have a knee-like structure.  
In Fig.~9a we also plot the prediction of the MCG simulation 
without the meson cloud obtained without fluctuations of the multiplicity
in $pp$ collisions. 
The ellipticity in this case has a weak knee-like structure at 
$c\sim 1 $\%. 
The knee for this version is seen better in Fig.~9b, that  
shows $\epsilon_2$ versus $N_{ch}$. To reduce the statistical fluctuations
the curves in Figs.~9a,b have been obtained by averaging of $\epsilon_2$
in the bins with the width $\Delta N_{ch}\sim 20$. For the two versions
(with and without the $MB$ component) with the fluctuating sources
the knee structure in Fig.~9b is absent. Thus we confirm the conclusion
of \cite{UU_Broniowski} that in the standard
MCG wounded nucleon scheme (without the $MB$ component) the fluctuations
erode the knee structure in $\epsilon_2$.   
This prediction is in contradiction with the analyses 
\cite{UU_Voloshin,UU_Singh} where the knee has been found.
From Fig. 9a one sees that at small centrality $\epsilon_2$ 
without fluctuations of the multiplicity in $pp$ collisions 
becomes smaller by
$20-10$\%.
This reduction is considerably bigger that the difference 
between our two versions with fluctuating sources.
Note that at $c\sim 0.01-1$\% the ratio of our prediction for $\epsilon_2$  
to the flow coefficient $v_2$ measured by STAR is 
$\sim 6-6.25$. It agrees qualitatively with the 
ratio $\epsilon_2/v_2$ obtained in hydrodynamical simulations
of Refs. \cite{Heinz_v2e2,Niemi_v2e2}.

An interesting way to investigate the mechanism of the entropy generation
in $AA$ collisions and the shape of the initial plasma fireball 
in collisions of nonsperical nuclei is the use for the centrality
categorization signals from 
the Zero-degree Calorimeters (ZDCs) \cite{Heinz_ZDC1507} 
that detect spectator neutrons.
Selecting the events with very low ZDC signals means selection  
of nearly full-overlap collisions with high multiplicity and very small 
number of the spectator neutrons \cite{UU_Heinz1,UU_STAR}.
The STAR collaboration \cite{UU_STAR} have measured the multiplicity
dependence of the flow coefficient $v_2$ for the top $1$\% and $0.1$\%
most central events selected on the smallnes of the ZDC signals in 
U+U collisions at $\sqrt{s}=193$ GeV and in Au+Au
collisions at $\sqrt{s}=200$ GeV. 
In the MCG wounded nucleon model the ZDC signal is 
usually mimiced by the number
of the spectator nucleons $N_s=2A-N_w$ \cite{UU_Heinz1,UU_Heinz}.
However, the equivalence of the centality categorization 
via the experimentaly measured ZDC activity and that via the $N_s$ in the MCG 
simulations is by no means evident. This is because physically it is clear
that the dynamical evolution of the hadron systems in the nucleus
fragmentation regions after the $AA$ collision is a complicated process 
that can involve interaction of the wounded and not wounded 
nucleons. These final state interactions, which are completeely 
ignored in the Glauber scheme, can reduce the number
of neutrons that could reach the ZDCs. For this reason the possibility to 
model the ZDCs event selection in terms of $N_s$ in the MCG 
simulations should rather be taken as a working hypothesis, 
to be explored in further studies.
 
In the present paper we ignore possible dynamical effects
in the nucleus fragmentation regions and following previous studies
\cite{UU_Heinz,UU_Bass} assume that the $N_s$ 
categorization reproduces
that through the ZDC signals. In terms of the $N_s$ the centrality
is defined as
\beq
c_s(N_s)=\sum_{N=0}^{N_s}P_s(N)\,,
\label{eq:280}
\eeq
where $P_s(N)$ is the probability distribution for $AA$ collisions
in $N_s$. The events with very small $c_s\ll 1$ correspond 
to collisions with a small impact parameter. For collisions of the 
nonspherical nuclei the initial asymmetry of the produced fireball 
is sensitive to angular orientation of the colliding nuclei. 
For an axis-symmetric
nucleus the orientation is described by the pair of the polar angles
$(\theta,\phi)$. The uranium nucleus has a prolate shape.
The high-overlap central tip-tip U+U collisions 
correspond to the polar angles with
$|\cos{\theta_1}|+|\cos{\theta_2}|\approx 2$.
In this case the azimuthal asymmetry
of the produced fireball should be dominated by the statistical 
fluctuations of the nucleon distributions, and should be small.
In the case of the highly overlapping body-body collisions  
with $|\cos{\theta_1}|+|\cos{\theta_2}|\ll 1$ and $|\phi_1-\phi_2|\ll \pi$
(or $|\phi_1-\phi_2|\approx\pi$) the fireball asymmetry should be larger due to the prolate shape of the
nucleus ellipsoids. In the MCG simulations
the highly overlapping collisions correspond to small values of $N_s$.
For this reason one can expect that for small $c_s$ the eccentricity
$\epsilon_2$ should decrease with charged multiplicity $N_{ch}$, because
the relative contribution of the $N_{coll}$ term to the entropy production
becomes bigger in the tip-tip collisions that give a smaller
ellipticity. For Au+Au collisions the situation
is opposite to that in U+U collisions, because the gold nucleus has an 
oblate form. 
For this reason in the two component Glauber scheme the charged 
multiplicity in the highly overlapping collisions should be smaller
in the tip-tip collisions, and one can expect that for small $c_s$ 
the eccentricity $\epsilon_2$ should grow with the charged multiplicity.  

In the upper panels of Fig. 10, we plot 
the linear fits of our  MCG results in the interval 
$0.9<N_{ch}/\langle N_{ch}\rangle<1.1$
for the rms $\epsilon_2$ versus the ratio $N_{ch}/\langle N_{ch}\rangle$ 
for the spectator centrality windows $c_{s}<0.1$\% (left) 
and  $c_{s}<1$\% (right) (the results are shown for the smearing width
$\sigma=0.7$ fm, but the results for $\sigma=0.4$ fm are similar).
One sees that, as expected, for the high-overlap collisions
in the window $c_{s}<0.1$\%  the slope of the curves is negative for U+U
collisions and positive for Au+Au collisions. The slope of the curves 
is flatter for the version with the $MB$ component, and for Au+Au collisions
this version gives pracically flat $\epsilon_2$.
For $c_{s}<1$\% the slope of the curves for U+U collisions become a bit lower,
but for Au+Au collisions the predictions are very close to that for 
$c_{s}<0.1$\%.
In the lower panels of Fig. 10 we compare the theoretical predictions
with the STAR data \cite{UU_STAR} on the flow coefficient $v_2$ 
for the top $0.1$\% and $1$\% ZDC centrality assuming that  
approximately $v_2\approx k\epsilon_2$ \cite{Heinz_v2e2,Niemi_v2e2}.
Due to the uncertainties in the value of the ratio $v_2/\epsilon_2$,
for each case  we simply  choose the values of $k$ 
to match the product $k\epsilon_2$ to the experimental $v_2$
at $N_{ch}/\langle N_{ch}\rangle\approx 1$.
One sees that there is a tendency that the reduction of the slope
in the version with the $MB$ component improves the agreement
with the data at the centrality window $<0.1$\%. For U+U collisions
in the $1$\% window the agreement with the data of the results 
without the $MB$ component is of similar quality to that with the $MB$ 
component. However, there is considerable disagreement with the data
for  Au+Au collisions. The experimental $v_2$ has considerable
negative slope. The theoretical curve for the version
without the $MB$ componet has a small positive slope. The slope
becomes positive for the version with the $MB$ componet, but it is much
smaller than that for the experimental data.
Thus, from Fig. 10 we see that the account of the $MB$ component
improves the agreement with the data (especially for U+U collisions) 
for the centrality window $c_s<0.1$\%. But for the window $c_s<1$\%
the situation is somewhat controversial and one cannot draw a definite
conclusion. It is possible that the problem for the window $c_s<1$\% is
due to inequivalence of the theoretical centrality categorization via $N_s$
and that via the ZDC signals used by STAR \cite{UU_STAR} that may become
stronger with increase of centrality.

\subsection{p+A and p+p collisions}
It is possible that a small size hot QGP may also be produced in $pA$ 
and even in $pp$ collisions. The idea that the QGP may be produced
in hadron collisions is very old \cite{Shuryak_QGP}.
The observations of the ridge effect in p+Pb collisions at $\sqrt{s}=5.02$ TeV 
\cite{CMS_ridge_pA,ALICE_ridge_pA,ATLAS_ridge_pA} and in high multiplicity 
$pp$ events at $\sqrt{s}=7$ TeV \cite{CMS_ridge_pp} support this idea.
The estimates from the observed charged multiplicities 
show that for the typical $pPb$ and $pp$ events at the LHC energies 
the initial temperature of the 
mini QGP  at the proper time $\tau\sim 0.5$ fm may be $\sim 250$ MeV
\cite{Z_mQGP1,Z_mQGP2},
which is well above the deconfinement temperature.
In the scenario with the mini QGP formation the ridge effect
in $pPb$ and $pp$ collisions may be connected with the hydrodynamic
expansion of the azimuthally asymmetric initial plasma fireball 
\cite{Bozek_pA_hydro,Bozek_pp_hydro,Romatschke_pp_hydro}.

In this subsection we present
our results of the MCG simulation of the mini fireball
in $pPb$ and $pp$ collisions at $\sqrt{s}=5.02$ TeV. 
As above for Pb+Pb collisions at $\sqrt{s}=5.02$ TeV 
for the parameter $\alpha$ we use the values obtained 
from the analysis of the centrality
dependence of the charged multiplicity density in Pb+Pb collisions
at $\sqrt{s}=2.76$ TeV.
In Fig. 11 we plot the results for the charged multiplicity density
$dN_{ch}/d\eta$ versus centrality for the versions with and without the
meson cloud. 
As for $AA$ collisions we define the theoretical centrality through the 
charged multiplicity distribution at $\eta=0$ (\ref{eq:180}).
We compare the results to available data from ALICE \cite{ALICE_pA828} obtained
via the centrality estimator corresponding to the central rapidity
region (CL1 in Fig. 16 of \cite{ALICE_pA828}). Note that the results
of \cite{ALICE_pA828} obtained via the centrality estimators 
with large $|\eta|$ intervals (V0M, V0A in Fig. 16 of \cite{ALICE_pA828})
and via the energy deposited in the neutron calorimeter on the Pb-going
side (ZNA in Fig. 16 of \cite{ALICE_pA828}) give somewhat weaker 
centrality dependence of the charged multiplicity density.
The fact that different centrality estimators give different results
are not surprising because for $pA$ collisions the fluctuations
of $N_{ch}$ at a given impact parameter are of the order of $N_{ch}$.
For this reason, contrary to $AA$ collisions where the role
of fluctuation turns out to be relatively small \cite{Broniowski_c}, 
the geometry of the $pA$ collision cannot be accurately determined 
from the observed charged multiplicity on the event-by-event basis. 
Despite the above uncertainties with definition of the centrality
for $pA$ collisions, Fig. 11 shows that at a qualitative level
the theoretical results agree with the data. A somewhat weaker decrease 
of the experimental charged multiplicity density with centrality may be
due to a considerably wider pseudorapidity region ($|\eta|<1.4$) used in 
\cite{ALICE_pA828} in the centrality categorization, while our procedure
corresponds to $|\eta|<0.5$. It is clear that for a broader $\eta$ region
the effect of the multiplicity fluctuations should be smaller, and 
the centrality categorization should be biased to the higher centralities.
From Fig. 11 one can see that the theoretical results obtained with 
and without the meson
cloud turn out to be very similar.
For the mean charged multiplicity density there is no problem with
the centrality selection. Our calculations give for the
whole centrality window
$dN_{ch}/d\eta\approx 19.5$,  which is in a reasonable agreement with 
the result $dN_{ch}/d\eta\approx 17.8$ for 
all centralities from Ref. \cite{ALICE_pA828}.
\begin{figure}%[ht]
\epsfig{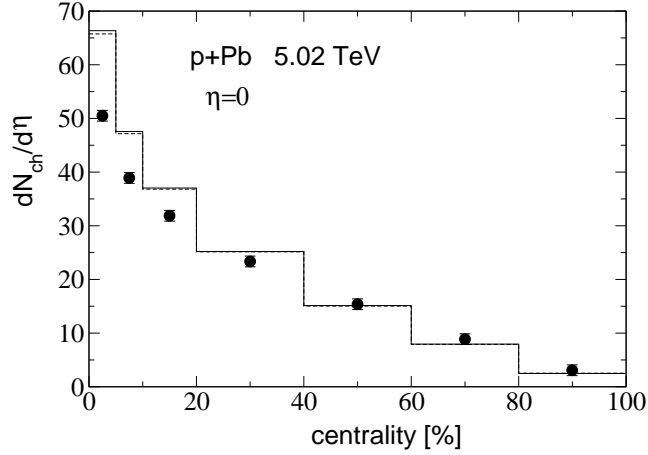}
\caption{\small Centrality dependence of $dN_{ch}/d\eta$ for 
p+Pb collisions at $\sqrt{s}=5.02$ TeV. Solid: 
MCG simulation for the scenario with the meson cloud
for $\alpha=0.09$. Dashed:
MCG simulation without the meson cloud
for $\alpha=0.14$. Data are from ALICE \cite{ALICE_pA828}.
}
\end{figure}

In Fig. 12 we show the results of the MCG simulations 
with and without the meson cloud for the event-averaged rms radius of 
the fireball versus the charged multiplicity 
density  for $pPb$ and $pp$
collisions at $\sqrt{s}=5.02$ TeV 
obtained for the smearing widths $\sigma=0.4$ and $0.7$ fm.
\begin{figure}
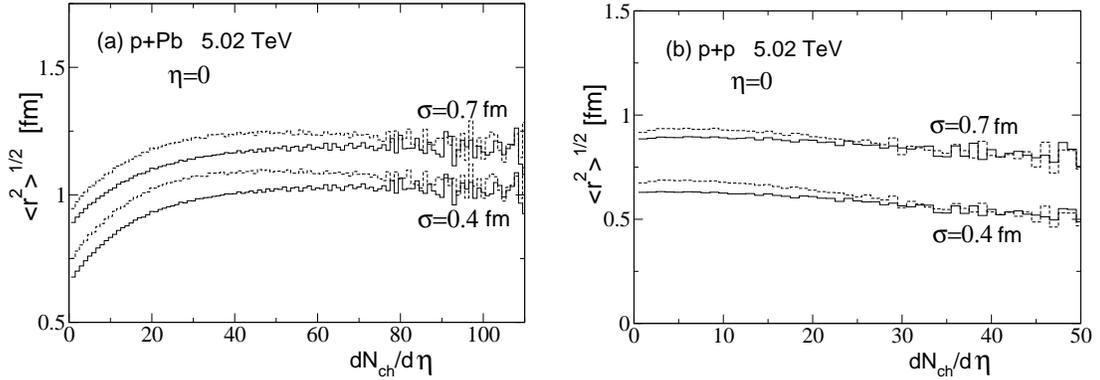
%[ht]
\epsfig{file=fig12a.eps,height=5cm,clip=}\hspace{.5cm}
\epsfig{file=fig12b.eps,height=5cm,clip=}
\caption{\small Multiplicity dependence of $\langle r^2\rangle^{1/2}$ for 
p+Pb (left) and p+p (right) collisions at $\sqrt{s}=5.02$ TeV
obtained from the MCG simulations for the Gaussian source width 
$\sigma=0.7$ and $0.4$ fm. Solid: 
the results for the scenario with the meson cloud
for $\alpha=0.09$.
Dashed: the results obtained without the meson cloud
for $\alpha=0.14$.
}
\end{figure}
One sees that for moderate charged multiplicities
the version with the $MB$ component gives somewhat smaller
fireball radius. It is due to smaller interaction radii in this
version. From Fig. 12 one can see that the increase of the fireball
radius with the smearing width $\sigma$ is less pronounced 
for $pPb$ collisions. This is because for $pPb$ collisions the geometry
of the fireball is largely controlled by the distribution 
of the nucleons in the nucleus around the path of the projectile proton.
The growth of the fireball size with multiplicity for $pPb$ collisions
is connected with the increase of the number of $NN$ interactions 
with large impact parameters.

\begin{figure}
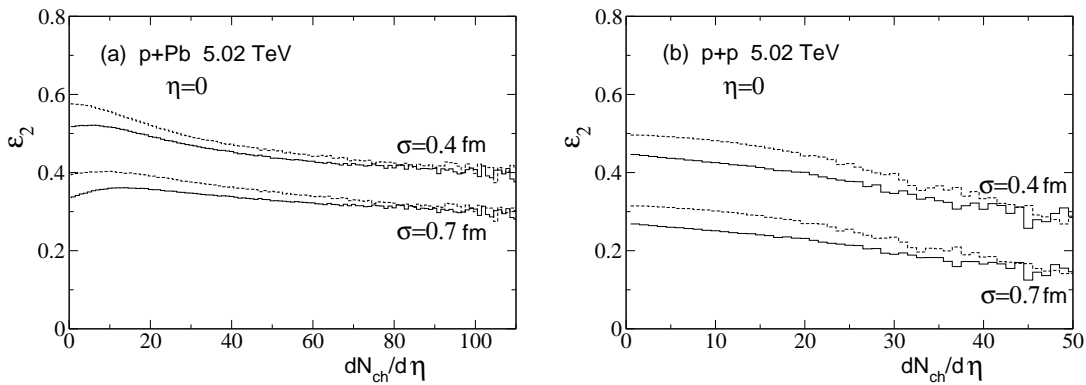
%[ht]
\epsfig{file=fig13a.eps,height=5cm,clip=}\hspace{.5cm}
\epsfig{file=fig13b.eps,height=5cm,clip=}\hspace{.5cm}
\caption{\small 
Multiplicity dependence of the rms $\epsilon_2$ for 
p+Pb (left) and p+p (right) collisions at $\sqrt{s}=5.02$ TeV
obtained from the MCG simulations for the Gaussian source width 
$\sigma=0.7$ and $0.4$ fm. Solid: 
the results for the scenario with the meson cloud
for $\alpha=0.09$.
Dashed: the results for the scenario without the meson cloud
for $\alpha=0.14$.
}
\end{figure}
In Fig. 13 we show the results for the multiplicity dependence 
of the ellipticity 
$\epsilon_2$ in $pPb$ and $pp$ collisions at $\sqrt{s}=5.02$ TeV. 
One sees that for both $pPb$ and $pp$ collisions the ellipticity
is smaller for the case with the $MB$ component.
The difference is more pronounced for $pp$ collisions.
The reduction of the ellipticity with the multiplicity for the version
without the $MB$ component is due to the growth of the fraction
of the hard component for high multiplicities. It is connected with
our choice of the position of the hard sources in the middle between
two colliding particles. This mechanism works for the version with the 
$MB$ component as well. Of course, the effect should be somewhat weaker 
due to a smaller value of $\alpha$. However, for the case with 
the $MB$ component there is an additional mechanism of the reduction of 
the ellipticity with multiplicity due to the events with simultaneous
inelastic interaction of the baryon and meson constituents in the $MB$
Fock component, that leads to a more symmetric fireball.
Note that our results for the rms $\epsilon_2$ in $pPb$ collisions 
for $\sigma=0.7$ fm are in a qualitative agreement with the rms flow 
coefficient $v_2\{2\}\sim 0.055-0.07$ obtained by CMS \cite{CMS_ridge_pA}
if one assumes that $\epsilon_2/v_2\sim 6$ as was obtained for
the large size QGP in the hydrodynamic simulations
of Refs. \cite{Heinz_v2e2,Niemi_v2e2}. 
Unfortunately, the elliptic flow $v_2$ from CMS \cite{CMS_ridge_pA}  is
given versus the number of tracks in the CMS detector, and one
cannot compare it directly with our $N_{ch}$ dependence in Fig.~13.
But anyway such a comparison could not be conclusive 
because our theoretical predictions for the geometry of the small size
fireball in $pA$ and $pp$ collisions depend crucially on the convention
for the transverse spacial positions of the entropy sources, and for
this reason they  are model dependent.

\section{Conclusions}
We have developed a Monte Carlo Glauber model for 
$AA$, $pA$ and $pp$ collisions that accounts for the $MB$
Fock component of the nucleon.
We have used the weight of the $MB$ component in the nucleon
wave function in the IMF that allows one to describe the DIS data
on the violation of the Gottfried sum rule \cite{ST}.
We have found that in the presence of the $MB$ Fock component
the required fraction of the binary collisions in the wounded nucleon
model becomes smaller.   
From the analysis of the STAR data \cite{STAR1} on the charged 
multiplicity density in Au+Au collisions at $\sqrt{s}=0.2$ TeV
we obtained the fraction of the binary collisions $\alpha=0.06$ and 
$0.135$ for the versions with and without the $MB$ component, respectively.
A similar fit to the  ALICE data \cite{ALICE1} 
on Pb+Pb collisions at $\sqrt{s}=0.2$ TeV 
gives for the two versions $\alpha=0.09$ and 
$0.14$.
Our results show that for central $AA$ collisions at the RHIC and LHC energies
the meson cloud can increase the multiplicity density in
the central rapidity region by $\sim 16-18$\%.
We have found that the $MB$ component leads 
to the growth of the ratio the charged multiplicity density to the number
of the wounded nucleons from peripheral to central
$AA$ collisions by $\sim 20$\% and $\sim 15$\% at RHIC
and LHC energies, respectively.

We have used the results of our fit to the data on the charged multiplicity 
density in Pb+Pb collisions at $\sqrt{s}=2.76$ TeV to give predictions
for the future LHC run 2 at $\sqrt{s}=5.02$ TeV. 
%From Fig. 6 one sees that 
As compared to $\sqrt{s}=2.76$ TeV we have found  
the growth of $dN_{ch}/d\eta$ in the central Pb+Pb collisions for 
$\sqrt{s}=5.02$ TeV by about $20$\%. 

We have found that the effect of the meson cloud on the eccentricity
$\epsilon_2$ in Au+Au and Pb+Pb collisions is relatively small, except for very
peripheral collisions,  where it reduces $\epsilon_2$ by $\sim 20$\%.
We have also studied the eccentricity $\epsilon_2$ for collisions
of the deformed uranium nuclei.  
For U+U collisions our MCG simulations with and without the $MB$
component give $\epsilon_2$, that has not a knee-like structure in
the top central collisions. 
Our results for $\epsilon_2$ for the version without the $MB$ 
component are in contradiction with the results of 
Refs. \cite{UU_Voloshin,UU_Singh} that predicted the knee-like structure
in $\epsilon_2$.
We find the knee structure only for the version of the MCG model 
without the meson cloud and without multiplicity fluctuations in 
$pp$ collisions. This is in agreement with the results of the analyses
\cite{UU_Lednicky,UU_Broniowski}.
We have also studied the multiplicity dependence of 
$\epsilon_2$ for the top central U+U and Au+Au collisions
with the centrality $c_s$ defined via the number of the spectator nucleons 
$N_s$ that is used to model the centrality categorization via the ZDC
signals \cite{Heinz_ZDC1507}. We have found that for $c_s<0.1$\%
the $MB$ Fock component improves the agreement with
the STAR data \cite{UU_STAR} on the $N_{ch}$ dependence of $\epsilon_2$.
But the results for $c_s<1$\% disagree with the data (especially for
Au+Au collisions). It is possible that it is 
due to inequivalence of the theoretical centrality categorization via $N_s$
and that via the ZDC signals used by STAR \cite{UU_STAR}.

We have also applied our MCG model to $pA$ and $pp$ collisions. 
We have found that the effect of the $MB$ component may be more important
for the initial asymmetry of the plasma fireball in $pA$ and $pp$ collisions,
where it gives reduction of the eccentricity $\epsilon_2$
by $\sim 15-20$\% for the typical charged multiplicities.
We have found that for the small size fireball produced in $pA$ and $pp$
collisions the MCG model with the meson cloud reduces the size of the
fireball by $10-20$\%. 

Because the $MB$ components are the long-range fluctuations in the physical
nucleon, one can expect that the observed effects should exist 
in other schemes of the entropy production. For this reason 
it would be of great interest to study the effect of the meson 
cloud within the IP-Glasma model \cite{IP-GL1,IP-GL2} 
(especially for $pA$ and $pp$ 
collisions, where the effect of the $MB$ component should be bigger).  

\begin{acknowledgments} 	
I thank 
W.~Broniowski and 
S.A.~Voloshin for communications.
This work is supported 
in part by the 
grant RFBR
15-02-00668-a.
\end{acknowledgments}

\end{document}